\documentclass[12pt, reqno]{amsart}
\usepackage{amsmath, amsthm, amscd, amsfonts, amssymb, graphicx, xcolor}
\usepackage[bookmarksnumbered, colorlinks, plainpages]{hyperref}

\theoremstyle{definition}

\theoremstyle{remark}

\numberwithin{equation}{section}

\begin{document}
\setcounter{page}{1}

\color{darkgray}
\noindent 
\hfill
\hfill

\centerline{}

\centerline{}

\title[Implementation of a Model of the Cortex Basal Ganglia Loop]{Implementation of a Model of the Cortex Basal Ganglia Loop}

\author[N. Arakawa]{Naoya Arakawa$^1$}

\address{$^{1}$ The Whole Brain Architecture Initiative}
\email{\textcolor[rgb]{0.00,0.00,0.84}{naoya.arakawa@nifty.com}}



\begin{abstract}
This article presents a simple model of the cortex-basal ganglia-thalamus loop, which is thought to serve for action selection and executions, and reports the results of its implementation. The model is based on the hypothesis that the cerebral cortex predicts actions, while the basal ganglia use reinforcement learning to decide whether to perform the actions predicted by the cortex.  The implementation is intended to be used as a component of models of the brain consisting of cortical regions or brain-inspired cognitive architectures.\end{abstract} \maketitle

\section{Introduction}

\noindent The cortico-basal ganglia-thalamo-cortical loop is thought to serve for action selection and execution \cite{OReilly2020}(Chapter 10).  Since the loop is common to most cortical areas, a model of the loop could serve as a component of models of the entire brain; it could serve not only for computational cognitive neuroscience, but also for the design of architectural parts for brain-inspired artificial general intelligence (AGI), as the cerebrum is involved in a variety of intellectual functions.
This article reports the implementation of a model based on the hypothesis that the cerebral cortex predicts actions, while the basal ganglia use reinforcement learning to decide whether to perform the actions predicted by the cortex (details follow).

\subsection{The Hypothesis}

The cerebral cortex selects/predicts actions.  The basal ganglia determine whether the cortex's selection is suitable for execution and disinhibit the thalamus to send a Go signal to the cortex at the appropriate timing.  
The basal ganglia (BG), as reinforcement learners, receive information on the state based on the input to the cortex and the action predicted, and decide whether to perform the action (Go/NoGo).  
The BG receive rewards and learn the State/Action \begin{math}\Rightarrow\end{math} Go/NoGo policy \cite{OReilly2020}.  The cortex learns from executed State/Action pairs and comes to predict (select) the appropriate State \begin{math}\Rightarrow\end{math} Action.  The hypothesis reconciles the role of reinforcement learning and prediction in cerebral action selection.

The following would support the hypothesis.
\begin{itemize}
 \item The cortex selects/predicts actions:
 \begin{itemize}
 \item The cortex is said to have predictive functions \cite{Millidge2021}.
 \item The prefrontal cortex is supposed to generate goal representation \cite{Passingham2012}.  Goals are thought to be generated predictively from inputs in the cortical areas.
 \end{itemize}
\end{itemize}
\begin{itemize}
 \item The BG control the timing of the start of an action (Go/NoGo) rather than the selection of the type of action:
 \begin{itemize}
 \item The BG are supposed to control the timing of motion initiation \cite{Dunovan2016}.
 \item As the matrix part of the thalamus, which is inhibited by the BG, projects to a relatively wide range of cortical areas \cite{Benarroch2008}\cite{Granger2007}, there may not be enough ``resolution" to distinguish between action options.
 \item The number of cells in the GPi of the BG, which controls the thalamus, is smaller than the corresponding number of minicolumns in the cortex (the number of minicolumns in the human neocortex is in the order of \begin{math}10^8\end{math} (\begin{math}10^{10}\end{math} neurons in the cortex \cite{Pakkenberg1997} / \begin{math}10^2\end{math} neurons in a minicolumn \cite{Sporn2005}), while the number of neurons in GPi is in the order of \begin{math}10^5\end{math} \cite{Thörner1975}) (It is arguable whether mini-columns code action choices).
 \end{itemize}
\end{itemize}

\subsection{Things Implemented}\leavevmode
\\
\\
\indent The following were implemented.
\begin{itemize}
 \item Winner-take-all (mechanism that selects from multiple actions)
 \item Separation of the cortex and BG
 \item Reinforcement learning in the BG
 \item Prediction learning in the cortex
\end{itemize}

\section{Implementation}

An overview of the implementation for the cortex, thalamus, and BG is described below.

\noindent The implementation code is at \url{https://github.com/rondelion/MinimalCtxBGA}.
\subsection{The Cortex}

It consists of the following parts.  The cortex receives observation and outputs action selection.
\begin{enumerate}
 \item \textbf{Output predictor}: Learns to predict actions from observation using the action ‘executed’ by the BG as the teacher signal.
A two-layer perceptron was used for implementation.
 \item \textbf{Output moderator}: Calculates the output from the output predictor and noise as input.  As the output predictior learns to make correct predictions, the contribution of noise decreases.
Specifically, the mean squared error was used as the probability for using output prediction.
 \item \textbf{Output selector}: The largest output from the output moderator is selected (winner-take-all) as one-hot-vector, and the output is gated with Go/NoGo signals from the BG.
\end{enumerate}
\subsection{The Thalamus}
The actual thalamus receives inhibitory signals from the BG (NoGo) and sends Go signals when disinhibited.  In the implementation, the Go/NoGo signal from the BG is directly transmitted to the output selector of the cortex, to omit the thalamus part (i.e., the thalamus was not implemented).
\subsection{The Basal Ganglia}
The BG were implemented as a monolithic reinforcement learner, where the concatenation of observation and action selection (one-hot-vector) in the output selector was used as the state, and reinforcement learning for Go/NoGo decisions was performed with reward from the environment.
\subsection{Overall Architecture}
Figure 1 shows the outline of the implementation with connections between the parts.
\begin{figure}[h]
  \centering
  \includegraphics[width=11cm]{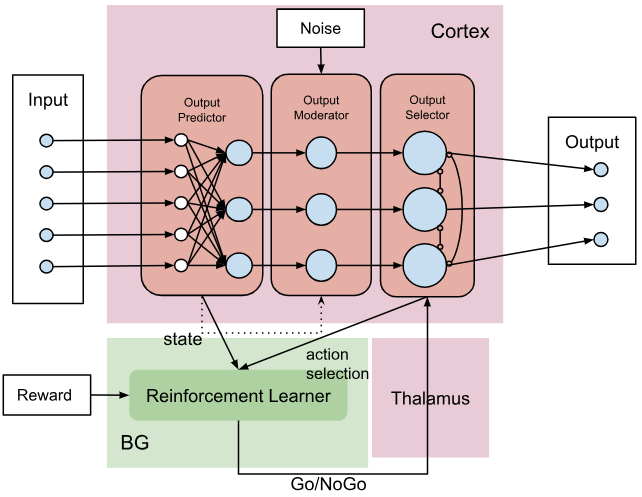}
  \caption{Overall Architecture}
\end{figure}
\section{Test Environment}
A simple delayed reward task in which a reward was given after action selection was implemented (CBT1Env.py).
\begin{itemize}
 \item Action selection is accepted while the cue is presented for a certain period of time.
A cue is one of the following: [1,1,0], [0,0,1], [0,1,1], [1,0,0]
Observation outside the presentation period is [0,0.0]
 \item If the relationship of the cue and action selection satisfies the following, reward will be given after a delay.
 \begin{itemize}
  \item Cues [1,1,0] or [0,0,1] with action [1,0]
  \item Cues [0,1,1] or [1,0,0] with action [0,1]
 \end{itemize}
 \item Penalty: If an output other than [0,0] is produced within the presentation period and the above reward conditions are not met, a negative reward will be given as a penalty (-0.7 in the experiment below).
If the delay before reward is greater than 1 unit time, the environment is no longer a Markov Decision Process and basic reinforcement learning algorithms cannot learn, so the experiment was conducted with a delay of 1 unit time.
\end{itemize}

\section{Implementation Details}
\subsection{Frameworks used}
\subsubsection{Cognitive architecture description framework}
The model was implemented as a module of BriCA (Brain-inspired Computing Architecture) \cite{Takahashi2015}, a computational platform for brain-inspired software development.  As mentioned in Introduction, the cortico-BG-thalamo-cortical loop is considered to be a generic component, so the model (module) would be incorporated as part of a larger cognitive architecture.  In the experiment, a cognitive architecture only consisting of the model was constructed and evaluated.
\subsubsection{Environment description framework}
OpenAI's Gym (the current version is \textit{Gymnasium} maintained by the Farama Foundation) was used.  Gym is widely used as a framework for agent learning environments.
\subsubsection{Machine learning framework}
PyTorch, a widely used machine learning framework, was used.  The perceptron was used for the cortical model, and reinforcement learning was used for the BG model.  For reinforcement learning, the DQN (Deep Q-Network) code from the PyTorch tutorial was used.  Though TensorForce, a reinforcement learning framework, was also tried, due to problems with its instability and not being maintained, the final version used the PyTorch DQN.
\subsection{Learning in the BG model}

The state value is the observation + output selector’s selected output concatenated.  The latter is the one-hot-vector before gating.  Its action is binary Go/NoGo to decide whether the entire loop is to take an action based on the state value.  The reward is from the external environment.  Action choice and its Go were carried out only once for each episode.  If no penalty was given, learning would not be successful, as Go would be opted for due to the positive reward.
Figure 2 shows the result of learning (average of 5 trials).
\begin{figure}[h]
  \centering
  \includegraphics[width=13cm]{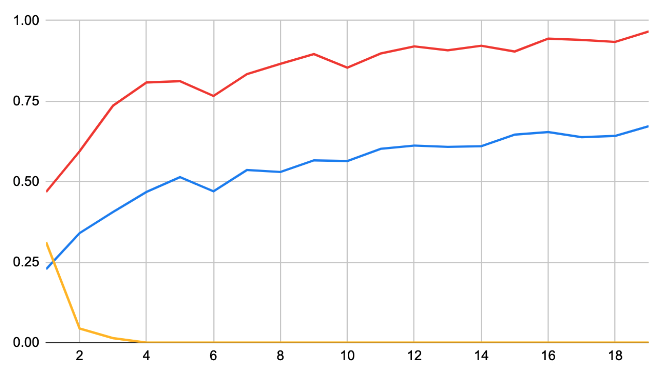}
  \caption{Learning Results\\
Horizontal axis: the number of episodes (×100)\\
Red line: average reward when Go is selected\\
Blue line: average reward\\
Yellow line: mean squared error of the cortical prediction
}
\end{figure}

\vspace{3mm}
\noindent \textbf{Comparison with Reinforcement Learning}

\noindent Figure 3 shows the results in which DQN (without the architecture being reported) was used to learn the above task (average of 5 trials).  As with the experiment above, cues were presented in one step and reward delay was one step.
\begin{figure}[h]
  \centering
  \includegraphics[width=10cm]{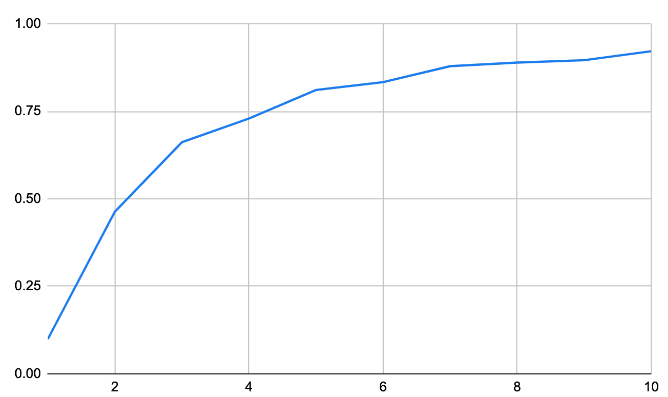}
  \caption{RL Results\\
Horizontal axis: the number of episodes (×100)\\
Vertical axis: average reward
}
\end{figure}

\subsection{Cortical Learning}
With external environmental input, the perceptron learned to predict action selection when the BG issued a Go.  The mean squared error was used as the probability that the prediction is used as the output candidate for which the BG part determines Go/NoGo (otherwise the output candidate is randomly determined).
Binary cross entropy (BCE -- which is said to be suitable for binary classification tasks) was used as the loss function of the perceptron.  Stochastic gradient descent (SGD) was used as the optimization algorithm, and its learning coefficient was set to 0.5 (through some trial and error).

The mean squared error is plotted with the yellow line in Figure 2.  The loss was decreasing as the BG learned.  Separate epoch loops (100 times per main episode) were run inside the cortex to augment the data (as the loss decreases quickly enough, it may need less augmentary loops).

\section{Discussion}
\subsection{Biological Plausibility}
The interior structure of the BG and thalamus was not implemented, and thus not subject to the plausibility evaluation.
While the implementation for the cortical model consists of three parts (output predictor, output moderator, and output selector), there is no evidence regarding the biological plausibility for this division.  Yet, the output of the cortex is at least related to Layer 5 of the cortex.  It is also known that cortical output is gated by Go/NoGo signals via the BG and thalamus \cite{Dunovan2016}\cite{OReilly2020}.  In the implementation, random selections of output values in the output moderator and the use of predictions in the output predictor were used as engineering hypotheses, to leave them as homework for neuroscience to search for functionally similar mechanisms.

If an action option corresponds to a cortical (mini-)column, lateral inhibition should be found between (mini-)columns to select the maximum value in a winner-takes-all way.  It may be realized as inhibition around columns \cite{Helmstaedter2009}.

As for connections between the cortex and the BG, the implemented cortical model passes the observation (or cortical state) and the output selected to the BG model.  The former corresponds to the connection from Layers III-Va to the striatal striosome, the latter to the connection from Layers Vb \& VI to the matrix in Fig.8 of \cite{Fujiyama2011}.

In the implementation, the cortex learns to predict executed action, and it could cause fixation on non-optimal solutions.  Possible countermeasures include always making random predictions with a certain probability and using rewards in learning.  The latter could be biologically plausible as dopamine signals are provided to the cortex \cite{Goldman-Rakic1997}.
While a reinforcement learner was used in the BG model, more biologically plausible implementations could be tried.  In particular, the DQN may not be biologically plausible as it uses tricks such as replay buffers.
\subsection{Related Topics}
Accumulation of evidence is claimed to occur in action decisions made around the BG \cite{Agarwal2018}\cite{Bogacz2007}.  The current implementation assumed that the input was reliable and it does not have an accumulation mechanism.  When dealing with issues involving uncertain input or multiple types of evidence, an accumulation mechanism should be introduced.
The Drift-Diffusion Model \cite{Dunovan2016}\cite{Ratcliff2008} to deal with the trade-off between action time pressure and situational understanding would also have to be implemented.

\subsection{Future Tasks}
The implementation was tested with a minimal delayed reward task.  Addressing more complex tasks  such as working memory tasks and tasks with non-Markov Decision Processes would require the introduction of short-term memory models in cortical regions, models of broader brain regions including hippocampal regions, and cortical control over those regions.  Wherever the cortical control is involved, the proposed model could be embedded to be evaluated.

\vspace{5mm}
\noindent {\bf Acknowledgement.} This work was partially financed by the Kakenhi project “Brain information dynamics underlying multi-area interconnectivity and parallel processing” in FY2021.

\newpage
\bibliographystyle{plain}

\end{document}